\documentclass[aps,prl,twocolumn,showpacs,amsmath,amsmath,amssymb,amsfonts]{revtex4-1}
\usepackage{graphicx}
\usepackage{dcolumn}
\usepackage{bm}
\usepackage{color}
\usepackage{multirow}

\begin{document}
\author{W. Li}
\affiliation{School of Physics and Astronomy, The University of Nottingham, Nottingham,
NG7 2RD, United Kingdom}
\author{C. Ates}
\affiliation{School of Physics and Astronomy, The University of Nottingham, Nottingham,
NG7 2RD, United Kingdom}
\author{I. Lesanovsky}
\affiliation{School of Physics and Astronomy, The University of Nottingham, Nottingham,
NG7 2RD, United Kingdom}
\title{Non-adiabatic motional effects and dissipative blockade for Rydberg atoms excited from optical lattices or microtraps}
\date{\today}
\begin{abstract}
The laser-excitation of Rydberg atoms in ultracold gases is often described assuming that the atomic motion is frozen during the excitation time. We show that this frozen gas approximation can break down for atoms that are held in optical lattices or microtraps. In particular, we show that the excitation dynamics is in general strongly affected by mechanical forces among the Rydberg atoms as well as the spread of the atomic wavepacket in the confining potential. This causes decoherence in the excitation dynamics - resulting in  a dissipative blockade effect - that renders the Rydberg excitation inefficient even in the anti-blockade regime. For a strongly off-resonant laser excitation - usually considered in the context of Rydberg dressing - these motional effects compromise the applicability of the Born-Oppenheimer approximation. In particular, our results indicate that they can lead to decoherence also in the dressing regime.
\end{abstract}

\maketitle

Ultracold laser-driven Rydberg gases are a versatile platform to study the coherent quantum dynamics in strongly interacting many-body systems. One reason that makes these systems so appealing is the fact that the thermal energy is so low that the atoms can be considered to be frozen in place on the timescale of laser-excitation \cite{anve+:98,moco+:98}. This absence of thermal atomic motion entails that the dynamics of ultracold Rydberg gases is entirely determined by the competition of the coherent laser-excitation process and the strong interaction between the highly excited atoms. This interplay results in an intricate and highly correlated excitation dynamics - its most prominent manifestation being the dipole-blockade: due to the strong interactions between Rydberg states only a single atom can be laser-excited within a certain exclusion volume, which in turn gives rise to an enhancement of the laser coupling to the emerging many-body state \cite{jaci+:00,lufl+:01}. Both, the dipole-blockade as well as the enhanced laser-coupling are at the heart of possible applications of ultracold Rydberg gases in quantum information science \cite{sawa+:10},  the simulation of quantum spin models \cite{welo+:08,olgo+:09,webu:10,le:11,jiat+:11,sepu+:11,zema+:12}, and the creation of highly nonlinear and nonlocal optical media \cite{prma+:10,sehe+:11,peot+:11,duku:12,pefi+:12}.

\begin{figure}
\includegraphics[width=\columnwidth]{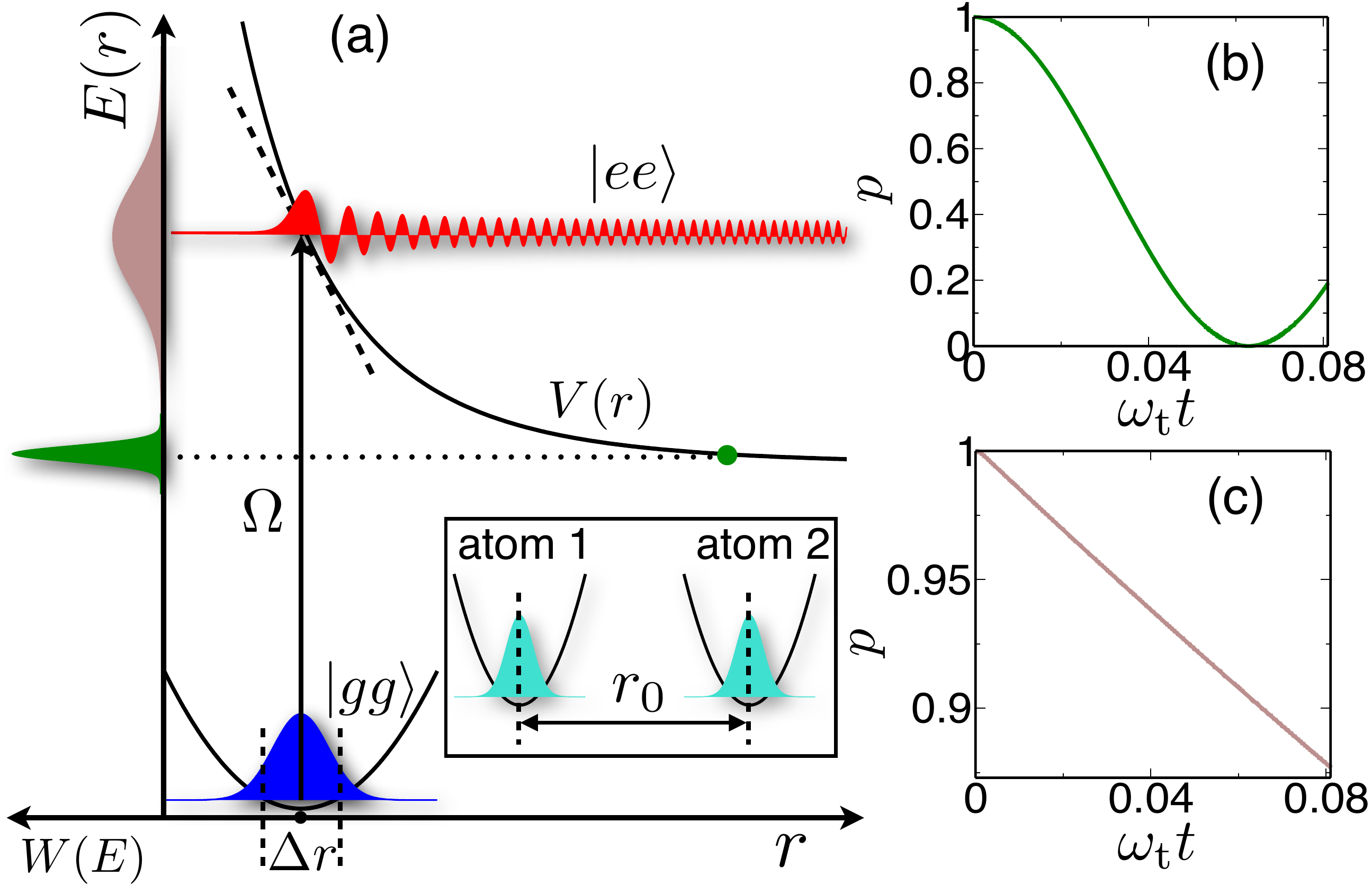}
\caption{(a) Anti-blockade configuration. Two atoms in their electronic ground state are prepared at a distance $r_0$ and subsequently resonantly photo-excited to a Rydberg pair state with Rabi frequency $\Omega$.  Each atom is initially assumed to be in the motional ground state of a harmonic potential (inset). The laser couples the corresponding wavefunction for the relative coordinate (blue) to a continuum of unbound Rydberg pair states (red). The overlap of these wavefunctions determines the energy-dependent laser coupling $W(E)$, which is shown for two different values of the initial separation $r_0$. For our analytical calculations we linearize the interaction potential around $r=r_0$ (dashed line). The numerically obtained time evolution of the survival probability $p(t)$ of the initial state is shown for the cases of large (b) and small (c) initial separation $r_0$ of the atoms taking into account the full potential $V(r) =C_6/r^6$ for the Rydberg $65S$ state of rubidium atoms with $C_6\approx 2\pi\times 370$ GHz$\cdot\mu\rm{m}^6$. The trapping potential is $\omega_{\rm{t}} = 2\pi\times 13.4$ kHz yielding $\Delta r\approx 133$ nm. The Rabi frequency is set to be $\Omega/\omega_{\rm{t}}=50$. The initial separation is $r_0/\Delta r=72$ in (b) and $r_0/\Delta r=30$ in (c). The time interval shown corresponds to $0.96\,\mu$s.
}
\label{fig:model}
\end{figure}

On a sufficiently long timescale dispersive forces among Rydberg atoms do eventually lead to atomic motion. However, this timescale is typically about one or two orders of magnitude larger than the excitation pulse duration. Due to this mismatch an adiabatic approach is often used where one considers the laser excitation in the frozen gas limit and treats the atomic motion separately within a molecular dynamics framework \cite{amre+:07a,wuat+:10,wuei+:11}. A similar route is also followed in the theoretical description of Rydberg dressing protocols \cite{hena+:10,pumi+:10,male+:10,howe+:10,wuat+:11,liha+:12,damu+:12}, where a strongly off-resonant laser coupling is used to admix a very small fraction of the Rydberg wavefunction to the atomic ground state. Here one conventionally applies a Born-Oppenheimer approximation, which consists of first determining effective potentials emerging from the off-resonant  laser coupling for fixed atomic position and then solving for the motional dynamics of the Rydberg dressed system.

In this work we show that for atoms which are trapped in optical lattices or microtraps the interplay between electronic and motional dynamics can already be important on typical timescales of resonant Rydberg excitation. To this end, we will investigate the quantum dynamics of a one dimensional model of the laser-excitation of two atoms, each trapped in a separate potential well. We will show that the excitation dynamics changes its character from fully coherent to dissipative when the distance between the atoms is decreased. The dissipative dynamics emerges when the mechanical force between the Rydberg atoms is so strong that the adiabatic approximation breaks down. This is accompanied with a significant slowing down of the excitation timescale leading to a dissipative blockade effect, where the laser excitation of a Rydberg atom pair gets inefficient even in the anti-blockade configuration. Moreover, our findings indicate this dissipation can also affect the dynamics of ground state atoms which are weakly dressed with a Rydberg state. In particular we will argue that the concept of an effective potential between the Rydberg dressed atoms can become insufficient to describe the dynamics of the system. Our results have implications on current and projected Rydberg experiments in optical lattices \cite{viba+:11,scch+:12} and microtrap setups \cite{gami+:09,urjo+:09,beve+:13p}.

In order to develop an understanding of the effect of atomic motion on the excitation process, we consider a model describing the laser excitation of a pair of atoms in one spatial dimension (cf. Fig.\ \ref{fig:model}). The electronic structure of each atom is modeled by two levels: the electronic ground state $\left| g \right>$ and an highly excited state $\left| e \right>$. These states are coupled by a laser field which is detuned by $\delta$ from the atomic transition and we parameterize the coupling strength by the Rabi frequency $\Omega_0$. We assume that two ground state atoms are initially prepared at a distance $r_0$, each in the lowest motional state of separate harmonic traps [see inset of Fig.\ \ref{fig:model}(a)]. This reflects the situation which is encountered in the case of atoms in a deep optical lattice or cooled to the motional ground state of microtraps \cite{kale+:12,thti+:13}. In this case the spatial wavefunction of each atom is a Gaussian of width $\sigma$. For $r_0 \gg \sigma$ the wavefunction describing the state of the relative motion of the atom pair is also a Gaussian $\chi(r;r_0) = (2\pi)^{-1/4} \Delta r^{-1/2} \exp\left[ -(r-r_0)^2/(4 \Delta r^2) \right]$ of width $\Delta r=\sqrt{2} \sigma$.

\emph{Anti-blockade configuration -}
Our first goal is to study the dynamics of the direct laser-excitation of a pair of Rydberg atoms as depicted in Fig.\ \ref{fig:model}(a). To achieve this the laser frequency needs to be chosen such that the anti-blockade condition $\Delta (r_0) \equiv 2 \delta + V(r_0) =0$ is met \cite{atpo+:07a,amgi+:10}, where $V(r)$ denotes the interaction potential of the Rydberg atoms. If $\left| V(r_0) \right|$ is much larger than the Rabi frequency $\Omega_0$ of the atomic transition, the anti-blockade condition also entails $|\delta| \gg \Omega_0$. In this regime the pair states with only one Rydberg excitation can be adiabatically eliminated and the Rabi frequency of the transition between the pair states $\left| gg \right>$ and $\left| ee \right>$ having zero and two Rydberg excitations, respectively, is $\Omega = \Omega_0^2/|\delta|$. These considerations, however, do not account for the spatial uncertainty of the initial pair state caused by the zero-point motion of the ground state atoms in their potential wells. Since the anti-blockade condition is strictly valid only for $r=r_0$, an uncertainty $\Delta r$ in the atomic separation will affect the laser excitation of the atom pair. In order to assess the consequences of this, we first investigate the dynamics of our model by solving it numerically.

The Hamiltonian of our system can be divided into a part describing the atomic motion and an electronic part accounting for the laser-excitation, $H = H_{\text{mot}} + H_{\text{el}}$, with ($\hbar=1$)
\begin{align}
H_{\text{mot}} &= - \frac{1}{2m} \nabla^2_r + U(r) \left| gg \right> \left< gg \right| + V(r) \left| ee \right> \left< ee \right| \label{eq:hmot} \\
H_{\text{el}} &= -\Delta (r) \left| ee \right> \left< ee \right| + \frac{\Omega}{2} \left( \left| gg \right> \left< ee \right| + \text{h.c.} \right) \, . \label{eq:hel}
\end{align}
Here, $m$ denotes the reduced mass of the atom pair and $U(r) = m \omega_t^2 (r-r_0)^2/2$ the harmonic trapping potential of the ground state atom pair with trapping frequency $\omega_t$. For simplicity we assume that the Rydberg atoms do not feel any external potential. In the following, we will focus on timescales that are much shorter than the radiative lifetime of the Rydberg atoms.

Panels (b) and (c) of Fig.\ \ref{fig:model} show the numerically obtained \footnote{We use a hybrid method combining the Crank-Nicolson algorithm to treat the atomic motion and a second order split operator method for the electronic coupling.} survival probability $p(t) = \left|\langle G | e^{-iHt} | G \rangle \right|^2$ of the initial state  $\left| G \right> = |gg\rangle \otimes \chi (r;r_0)$ as a function of time for large and small initial separation $r_0$ of the atoms, respectively. Evidently, the dynamics in these two regimes has a strikingly different character. While for a large separation (b) the initial state gets depopulated quickly and coherently, panel (c) displays an exponential decay of $p(t)$. The major difference of these two regimes is the variation of the interaction potential $V(r)$ over the spread $\Delta r$ of the relative wavefunction. This implies that the laser actually does not just couple two discrete electronic states ($\left| gg \right> \leftrightarrow \left| ee \right>$) but instead the discrete state $| G \rangle$ and a continuum of states $\left| E \right> = |ee\rangle \otimes \phi(r,E)$ of energy $E$ within an energy window $\sim |F| \Delta r$. Here $F= -\left. \partial_r V(r) \right|_{r=r_0}$ is the force between the Rydberg atoms at distance $r_0$. For a weak force the laser coupling is essentially constant over $\Delta r$ and only continuum states within a very small energy window are involved in the excitation dynamics. In the case of a strong force, however, $V(r)$ varies significantly over $\Delta r$ so that the laser coupling is smeared out over a large energy interval.

In order to obtain an analytical understanding of the excitation dynamics we employ the framework of Fano theory \cite{fa:61,bara:05}. We will assume in the following that the oscillator ground state is energetically well isolated from higher oscillator levels \footnote{Note that the approach can be generalized to the case where more oscillator levels are initially populated \cite{bara:05}.}. In this regime the Hamiltonian of our model takes on the form
$
H = \Delta(r_0) | G \rangle \langle G | +  \int \text{d} E \, E | E \rangle \langle E | + \int \text{d} E \, W(E) \left(| E \rangle \langle G | + \text{h.c.} \right)
$.
The energy-dependent coupling between the discrete state and the continuum, denoted by $W(E)$, is proportional to the spatial overlap between $\chi (r;r_0)$ with the continuum state $\phi(r,E)$ and given by $W(E) = (\Omega/2) \int_0^{\infty} \text{d}r \chi(r;r_0) \phi(r,E)$. The eigenstates of $H$ can be expressed as $\left| \omega \right> = \alpha (\omega) \left| G \right> + \int_{-\infty}^{\infty} \text{d}E \, \beta (\omega,E ) \left| E \right>$ with coefficients $\alpha (\omega)$ and $\beta (\omega,E)$ that have to be determined self-consistently \cite{bara:05}. Choosing the initial state to be $|G\rangle$ the probability to remain in it at time $t$ then is
$
p(t) = \left| \int_{-\infty}^{\infty} \text{d}\omega \, \left| \alpha (\omega) \right|^2 e^{-i\omega t} \right|^2 \, .
$
The dynamics of the system is therefore encoded in the Fourier transform of the spectral function,
$
\left| \alpha (\omega) \right|^2 = W^2(\omega)/\left\{ [\omega -\Delta(r_0) - \epsilon (\omega) ]^2 + \pi^2 W^4 (\omega) \right\}
$,
with level-shift function $\epsilon (\omega) = \mathcal{P}\int_{-\infty}^{\infty} \text{d}E\, W^2(E) / (\omega - E)$, where $\mathcal{P}$ denotes the principal value integral. The spectral function is normalized such that $\int_{-\infty}^{\infty} \text{d}\omega \, \left| \alpha (\omega) \right|^2 =1 $.

\begin{figure}
\includegraphics[width=\columnwidth]{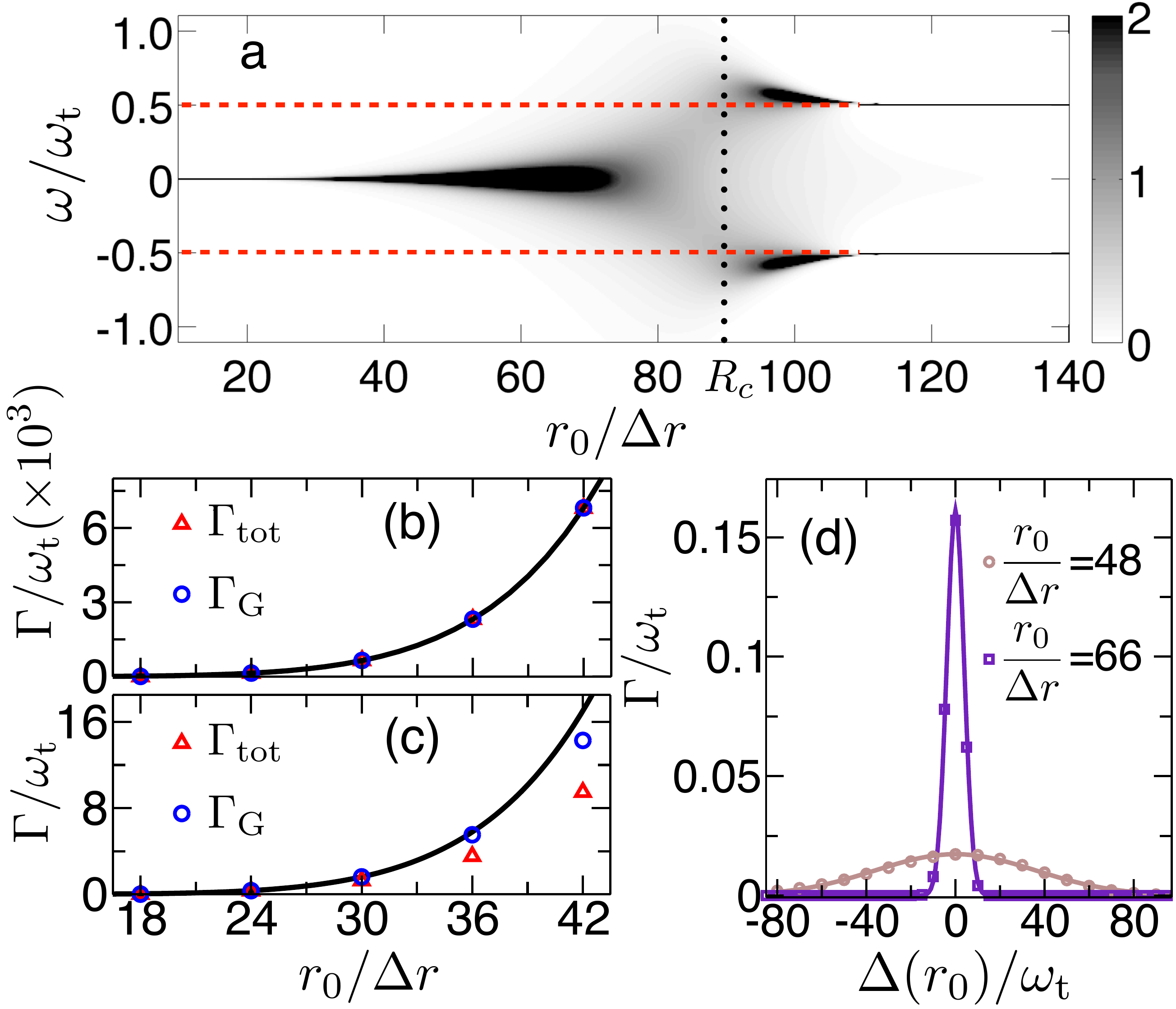}
\caption{(a) Dependence of the spectral function $\left|\alpha (\omega) \right|^2$ on the initial separation $r_0$ of the atom wavepackets for $\Delta (r_0) =0$ (see Fig. \ref{fig:model}) and $\Omega/\omega_{\rm{t}}=1$. The dashed lines indicate the positions of the eigenstates of the model when motional effects are not taken into account. The distance $R_c$, where the spectrum changes from a single to a double peak structure is indicated by the vertical dotted line. Panels (b) and (c): Decay rate in the strong force regime as function of the initial atom separation for $\Omega/\omega_{\rm{t}}=1$ (b) and $\Omega/\omega_{\rm{t}}=50$ (c). Comparison of the numerical solution of the model (symbols) and the analytical prediction obtained with Fano theory (solid line). (d) Decay rate in the strong force regime as function of $\Delta (r_0)$ for two different values of the initial atomic separation and $\Omega/\omega_{\rm{t}}=1$. The symbols and solid lines are the numerical solution and the predictions of Fano theory, respectively.
Numerical parameters that are not explicitly indicated in the panels are those of Fig.\ \ref{fig:model}.
}
\label{fig:fano}
\end{figure}

To determine the spectral function analytically, we make two approximations: (i) We linearize the interaction potential around $r_0$, i.e., we approximate $V(r) \approx V(r_0) + F (r-r_0)$.  The continuum states $\phi(r,E)$ are then the eigenstates of a particle in a linear potential, $\phi(r,E) = \mathcal{N}^{-1/2} \Phi \left\{ -\left[ (r-r_0) + E/F \right] /l_0  \right\}$, where $\Phi(x)$ denotes the Airy function \cite{lali_3:77}. The constant $\mathcal{N} = \pi |F|^{1/3}/(2m)^2$ is chosen such that these eigenfunctions are normalized to a $\delta$-function in energy and $l_0 = \left[ 1/(2 m |F|)  \right]^{1/3}$. (ii) We consider the regime where $\Delta r/l_0 \gg 1$. Within these approximations the coupling acquires a simple form,
$
W(E) = \frac{\Omega}{2 \pi^{1/4} \Delta E^{1/2}} \exp\left(- \frac{E^2}{2 \Delta E^2} \right)
$,
with energy width $\Delta E = \sqrt{2} \Delta r |F|$. In this Gaussian coupling regime the level shift function can also be expressed analytically,
$
\epsilon (\omega) = \frac{\Omega^2}{2 \Delta E} \mathcal{D}\left( \frac{\omega}{\Delta E} \right)
$,
where $\mathcal{D}(x) = e^{-x^2} \int_0^x \text{d}y \, e^{y^2}$ is the Dawson function.

The spectral function $\left| \alpha (\omega) \right|^2$ in the anti-blockade configuration [$\Delta (r_0)=0$] is shown in Fig.\ \ref{fig:fano}(a) as a function of the initial distance of the atomic pair for $\Omega/\omega_{\rm{t}}=1$. It displays a transition from a single peak to a double peak structure at a critical distance $R_c = 1.363 [V(\Delta r)/\Omega)]^{1/7} \Delta r$ assuming a van der Waals interaction, $V(r)=C_6/r^6$, between the Rydberg atoms. In this transition region the spectral function is broad, while for small/large $r_0$, i.e., strong/weak force, its peaks are very narrow. For $r_0/\Delta r \gg 1$ the peaks are located at $\omega = \pm \Omega/2$. This is the result one would obtain for the anti-blockade irrespective of the atomic separation if one completely neglected the variation of the interaction potential over the spatial width of the initial wavefunction. In this case, which corresponds to panel (b) in Fig.\ \ref{fig:model}, the spectral function is $\left| \alpha (\omega) \right|^2 = \left[ \delta \left( \omega + \Omega/2 \right) + \delta \left(\omega - \Omega/2 \right) \right]/2$. It has two $\delta$-peaks at the eigenenergies of the electronic Hamiltonian (\ref{eq:hel}) which are the dressed energies of the fully coherent system [depicted also as dashed lines in Fig.\ \ref{fig:fano}(a)]. The change of the spectral function with the atom separation for finite $\Delta r$ therefore clearly shows that motional effects can have a significant effect on the excitation dynamics when the force between Rydberg atoms is strong.

To study this further, we focus on the regime, where $r_0 \ll R_c$. Here, the coupling $W(E)$ is spread over a very large energy interval $\Delta E$, such that it can be considered  constant as a first approximation. The spectral function in this regime acquires a Breit-Wigner form,
\begin{equation}
\left| \alpha (\omega) \right|^2 = \frac{1}{\pi} \frac{\Gamma/2}{(\Gamma/2)^2 + \{ \omega - \Delta(r_0) - \epsilon[\Delta(r_0)]\}^2}
\end{equation}
i.e., the survival probability of the initial state decays exponentially, $p(t) = \exp\left( -\Gamma t\right)$, as shown in Fig.\ \ref{fig:model}(c) with rate
\begin{equation}
\Gamma = \frac{\sqrt{\pi}}{2} \frac{\Omega^2}{\Delta E} \exp \left(- \frac{\Delta^2(r_0)}{\Delta E^2} \right) \, .
\label{eq:rate}
\end{equation}

Let us compare these predictions with results obtained by numerically solving our model and first focus on the regime, where the assumption of an energetically well isolated initial state is well met. Fig.\ \ref{fig:fano}(b) shows a comparison of $\Gamma$ with the decay constant extracted from fully numerical simulations as a function of the initial atomic distance for $\Delta (r_0) =0$. The numerical data were obtained by analyzing the time evolution of the survival probability of the initial state ($\Gamma_G$) and by following the dynamics of the total population in the harmonic potential ($\Gamma_{\text{tot}}$). The analytical predictions are in excellent agreement with the numerical data.  In addition, also for finite detuning the agreement between the numerical simulations and the analytical prediction of Eq.\ (\ref{eq:rate}) is very good.  This can be seen in Fig.\ \ref{fig:fano}(d) showing the rate as function of $\Delta(r_0)$ for two different initial separations. The slight asymmetry of the numerical data about $\Delta (r_0) = 0$ stems from the fact that the force between the Rydberg atoms is not constant over the initial wavepacket as assumed in our analytical calculations. Interestingly, as demonstrated in Fig.\ \ref{fig:fano}(c), our analytical analysis also gives reasonable agreement when the assumption of an energetically isolated initial state does not hold.  At small spatial separations the agreement between numerical and analytical results is remarkably good. For larger $r_0/\Delta r$ the full solution shows that the oscillator ground state gets coupled to higher levels via the continuum. However, irrespective of this, the excitation dynamics dramatically slows down with increasing mechanical force between the Rydberg atoms and the excitation rate is inversely proportional to it. Thus, although on resonance, the Rydberg excitation becomes more and more inefficient with increasing mechanical force. This effect can be viewed as a dissipative excitation blockade induced by decoherence due to atomic motion.

\begin{figure}
\centering
\includegraphics[width=0.85\columnwidth]{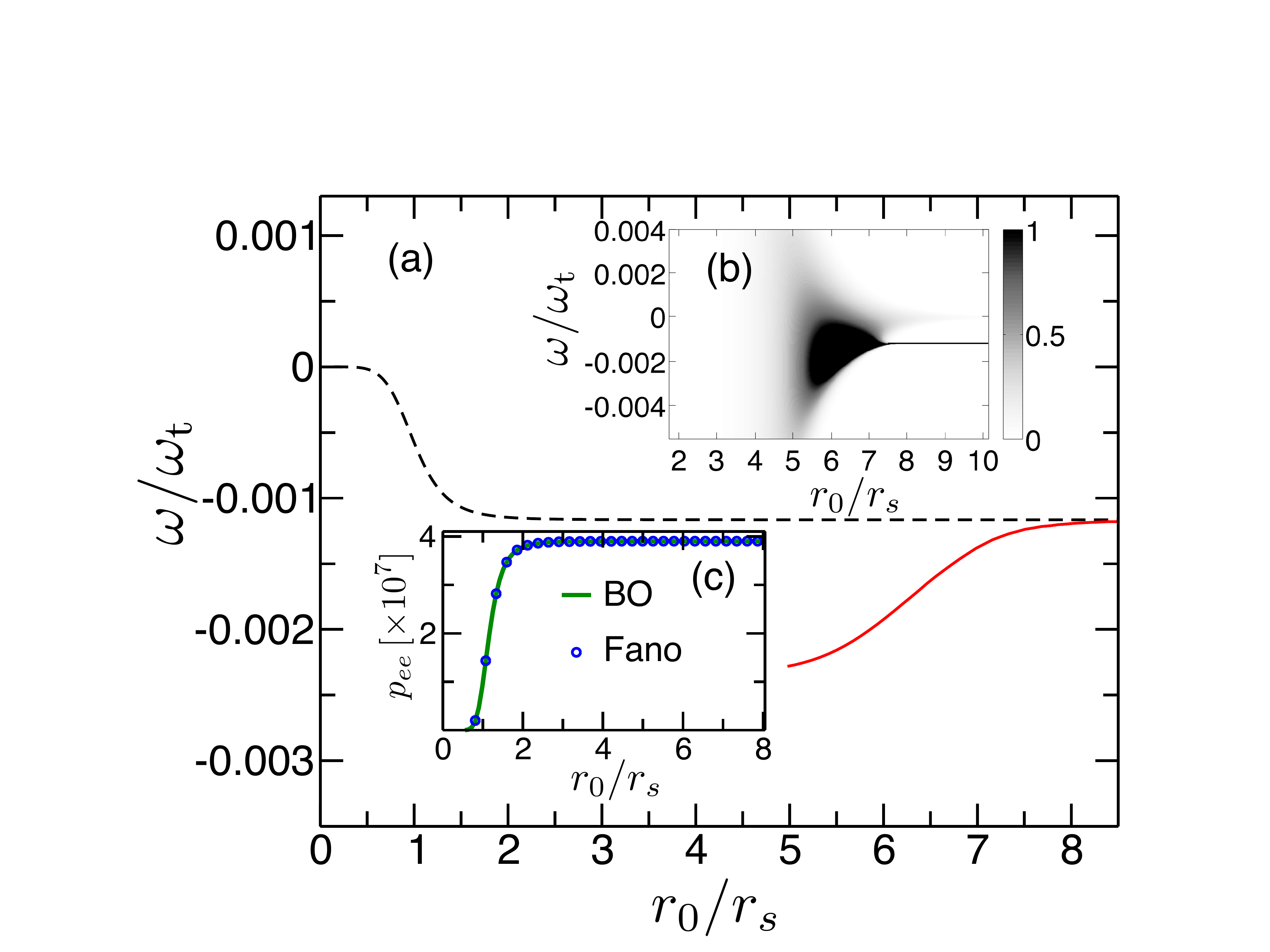}
\caption{(a) Distance dependence of the energy (black dashed line) of a Rydberg dressed atom pair $\omega_D (r_0)$ in the motional state $\chi (r,r_0)$. (b) The spectral function $\left| \alpha (\omega) \right|^2$ of the system showing a significant spectral broadening as the soft-core radius is approached. The peak position of $\left| \alpha (\omega) \right|^2$ is shown as solid red line in panel (a) up to the point where the width of the spectral peak starts to exceed its height. (c) Distance dependence of the probability to be in the Rydberg pair state for $t\to\infty$. The data set obtained from the Fano theory is given by the circles while the solid line shows the results from the Born-Oppenheimer  approximation. The numerical parameters used are $\delta = 2\pi\times 20$ MHz,  $\Omega_0=2\pi\times 1$ MHz and $r_{\rm{s}}/\Delta r \approx 34.5$.
}
\label{fig:dressing}
\end{figure}

\emph{Dressing regime -}
Let us finally analyze the situation, in which the excitation laser is far-detuned from the single atom transition as well as from the anti-blockade condition. This is the regime of Rydberg dressing, where only a very small fraction of the Rydberg wavefunction is admixed to the ground state atoms \cite{hena+:10,pumi+:10,liha+:12}. The main effect of this small admixture is that a pair of Rydberg dressed atoms exhibits a distance-dependent energy shift. In the Born-Oppenheimer approximation, the spectral function consists of two $\delta$-peaks at the eigenenergies of the $r$-dependent electronic Hamiltonian - as in the anti-blockade configuration. The lowest eigenenergy corresponds to the dressing potential and is given by $\tilde{\omega}(r) = \left[\Delta (r) - \sqrt{\Delta^2(r) + \Omega^2} \right]/2$. The black dashed line in Fig.\ \ref{fig:dressing}(a) depicts the interaction energy of two dressed atoms in the motional state $\chi(r;r_0)$ which is given by the convolution $\omega_D (r_0) = \int \text{d}r\, |\chi(r;r_0)|^2 \tilde{\omega}(r)$. For a van der Waals interaction this energy exhibits a characteristic soft-core with radius $r_{s}=(C_6/2 |\delta|)^{1/6}$ and height $V_0 = \Omega_0^4/8|\delta|^3$. Moreover, the probability to be in the Rydberg pair state is given by, $p_{ee}(r_0)= \int\text{d}r\, | c_{ee}(r,r_0) |^2$, where $c_{ee} (r,r_0)$ is the coefficient of the Rydberg pair state in the eigenstate $| \tilde{\omega}(r) \rangle = c_{gg} (r,r_0) |gg \rangle + c_{ee} (r,r_0) |ee\rangle$ corresponding to the eigenvalue $\tilde{\omega}(r)$. This probability is shown as a function of $r_0$ as solid line in Fig.\ \ref{fig:dressing}(c).

This picture is changed when the effects of atomic motion are included in the description. Fig.\ \ref{fig:dressing}(b) shows the spectral function $|\alpha (\omega)|^2$ obtained from Fano theory. For $r_0\to\infty$ one recovers the $\delta$-peak at the eigenenergy of the dressed system. However, as the soft-core radius $r_s$ is approached this peak broadens and the spectral weight is spread over an energy interval largely exceeding $V_0$. Eventually the broadening is so large that the spectral function is almost flat and no pronounced structure is visible in $|\alpha (\omega)|^2$. Moreover, as shown by the solid red line in Fig.\ \ref{fig:dressing}(a) the peak position is shifted as compared to the adiabatic energy of the dressed state. We show the data up to the point where the peak height becomes smaller than its width. The large broadening of the spectral function entails that the energy of the Rydberg dressed state gets less and less well defined with decreasing atomic separation. This suggests the emergence of a dissipative dynamics for sufficiently small interatomic distances. Comparing the excitation probability to be in the Rydberg pair state in the limit $t\to\infty$, $p_{ee} (r_0) = \int_{\omega_{\text{min}}}^{\omega_{\text{max}}} \text{d} \omega \, |\alpha (\omega)|^2$ (the integration limits ($\omega_{\text{min}/\text{max}}$) are chosen such that they contain the entire peak), we find that this probability actually coincides with the results of the Born-Oppenheimer approximation [see Fig.\,\ref{fig:dressing}(c)].

Neglecting the effect of atomic motion, one thus obtains the correct probability to be in the electronic states $|gg\rangle$ or $|ee\rangle$, however, without further information about the character of the excitation dynamics. This illustrates that the inclusion of motional effects in the description of the excitation process can be important even in the far off-resonant regime of Rydberg dressing.

\begin{acknowledgments}
\emph{Acknowledgements --- }
We acknowledge funding from EPSRC and the ERA-NET CHIST-ERA (R-ION consortium). C.A. acknowledges support through a Feodor-Lynen Fellowship of the Alexander von Humboldt Foundation.
\end{acknowledgments}

\bibstyle{revtex4-1}

\end{document}